\documentclass[aps,prd,twocolumn,superscriptaddress,floatfix,showpacs,showkeys,preprintnumbers]{revtex4}
\usepackage[utf8]{inputenc}
\usepackage[T1]{fontenc}
\usepackage{slashed}
\usepackage{times}
\usepackage{amssymb,amsfonts,amsmath,amsthm}
\usepackage{dsfont,bbm}
\usepackage{dcolumn}
\usepackage{epsf}
\usepackage{graphicx}
\usepackage[caption=false]{subfig}
\usepackage{dsfont}
\usepackage{slashed}
\usepackage[active]{srcltx}
\usepackage[usenames]{color}

\begin{document}

\title{Constraining gluon poles}

\author{I.~V.~Anikin}
\email{anikin@theor.jinr.ru}
\affiliation{Bogoliubov Laboratory of Theoretical Physics, JINR,
             141980 Dubna, Russia}
\affiliation{Institut f\"ur Theoretische Physik, Universit\"at
Regensburg,D-93040 Regensburg, Germany}

\author{O.~V.~Teryaev}
\email{teryaev@theor.jinr.ru}
\affiliation{Bogoliubov Laboratory of Theoretical Physics, JINR,
            141980 Dubna, Russia}

\begin{abstract}
In this letter, we revise the QED gauge invariance for the hadron tensor
of Drell-Yan type processes with the transversely polarized hadron.
We perform our analysis within the Feynman gauge for gluons and
make a comparison with
the results obtained within the light-cone gauge.
We demonstrate that QED gauge invariance leads, first,
to the need of a non-standard diagram
and, second, to the absence of gluon poles in the
correlators $\langle\bar\psi\gamma_\perp A^+\psi\rangle$
related traditionally to $dT(x,x)/dx$.
As a result, these terms disappear from the final
QED gauge invariant hadron tensor.
We also verify the absence of such poles by analysing the
corresponding light-cone Dirac algebra.
\end{abstract}
\date{\today}
\maketitle

\section{Introduction}

In the recent times, we have observed the renaissance in
the nucleon structure studies through the Drell-Yan type processes
in the existing (FermiLab, Relativistic
Heavy Ion Collider, see  \cite{RHIC,Gamberg:2013kla}) and future (J-Parc, NICA) experiments.
One of the most interesting subjects of such experimental studies in this direction
is the so-called single spin asymmetry (SSA) which is expressed with
the help of the hadron tensor, see for instance \cite{Qiu:1991pp} or
\cite{Teryaev, Boer}.

Lately, we have reconsidered \cite{AT-GP} this process in the contour gauge.
We have found that there is a contribution from the {\it non-standard} diagram which
produces the imaginary phase required to have the SSA. This additional contribution
leads to an extra factor of $2$ for the asymmetry. This
conclusion was supported by analysis of the QED gauge invariance of the hadron tensor.

In comparison, the analysis presented in \cite{Zhou:2010ui} which uses the axial and Feynman gauges
does not support the latter conclusion. For this reason, we perform here the
detailed analysis of hadron tensor in the Feynman gauge with the particular emphasis
on the QED gauge invariance.
We find that the QED gauge invariance can be maintained only by taking into account the non-standard diagram.
Moreover, the results in the Feynman and contour gauges coincide if the
gluon poles in the correlators $\langle\bar\psi\gamma_\perp A^+\psi\rangle$ are absent.
This is in agreement with the relation between gluon poles and the Sivers function which
corresponds to the "leading twist" Dirac matrix $\gamma^+$.
We confirm this important property by comparing the light-cone dynamics for different correlators.

As a result, we derive the QED gauge invariant hadron tensor which
completely coincides with the expression obtained within
the light-cone contour gauge for gluons, see \cite{AT-GP}.

\section{Kinematics}

We study the hadron tensor which contributes to the single spin
(left-right) asymmetry
measured in the Drell-Yan process with the transversely polarized nucleon (see Fig.~\ref{Fig-DY}):
\begin{eqnarray}
N^{(\uparrow\downarrow)}(p_1) + N(p_2) &\to& \gamma^*(q) + X(P_X)
\nonumber\\
&\to&\ell(l_1) + \bar\ell(l_2) + X(P_X).
\end{eqnarray}
Here, the virtual photon producing the lepton pair ($l_1+l_2=q$) has a large mass squared
($q^2=Q^2$)
while the transverse momenta are small and integrated out.
The left-right asymmetry means that the transverse momenta
of the leptons are correlated with the direction
$\textbf{S}\times \textbf{e}_z$ where $S_\mu$ implies the
transverse polarization vector of the nucleon while $\textbf{e}_z$ is a beam direction \cite{Barone}.

Since we perform our calculations within a {\it collinear} factorization,
it is convenient to fix  the dominant light-cone directions as
\begin{eqnarray}
\label{kin-DY}
&&p_1\approx \frac{Q}{x_B \sqrt{2}}\, n^*\, , \quad p_2\approx \frac{Q}{y_B \sqrt{2}}\, n,
\\
&&n^{*\,\mu}=(1/\sqrt{2},\,{\bf 0}_T,\,1/\sqrt{2}), \quad n^{\mu}=(1/\sqrt{2},\,{\bf 0}_T,\,-1/\sqrt{2}).
\nonumber
\end{eqnarray}
So, the hadron momenta $p_1$ and $p_2$ have the plus and minus dominant light-cone
components, respectively. Accordingly, the quark and gluon momenta $k_1$ and $\ell$ lie
along the plus direction while the antiquark momentum $k_2$ -- along the minus direction.
The photon momentum reads (see Fig.~\ref{Fig-DY})
\begin{eqnarray}
q= l_1+l_2=k_1 + k_2\,
\end{eqnarray}
which, after factorization, will take the form:
\begin{eqnarray}
q= x_1 p_1 + y p_2\,+ q_T.
\end{eqnarray}

\section{The DY hadron tensor}

We work within the Feynman gauge for gluons.
The standard hadron tensor
generated by the diagram depicted in Fig.~\ref{Fig-DY}(the left panel) reads
\begin{eqnarray}
\label{HadTen1-2}
&&d{\cal W}_{(\text{Stand.})}^{\mu\nu}=\int d^4 k_1\, d^4 k_2 \, \delta^{(4)}(k_1+k_2-q)\times
\nonumber\\
&&
\int d^4 \ell \,
\Phi^{(A)\,[\gamma_\beta]}_\alpha (k_1,\ell) \, \bar\Phi^{[\gamma^-]} (k_2)\times
\nonumber\\
&&
\text{tr}\big[
\gamma^\mu  \gamma^\beta \gamma^\nu \gamma^+ \gamma^\alpha
S(\ell-k_2)
\big]\, ,
\end{eqnarray}
where
\begin{eqnarray}
\label{PhiF}
&&\Phi^{(A)\,[\gamma_\beta]}_\alpha (k_1,\ell)
=
\\
&&{\cal F}_2\Big[
\langle p_1, S^T | \bar\psi(\eta_1)\gamma_\beta  gA_{\alpha}(z)  \psi(0) | S^T, p_1\rangle \Big] ,
\nonumber\\
&&\bar\Phi^{[\gamma^-]}(k_2)={\cal F}_1 \Big[
\langle p_2 | \bar\psi(\eta_2)\gamma^- \psi(0)| p_2\rangle \Big].
\end{eqnarray}
Throughout this paper, ${\cal F}_1$ and  ${\cal F}_2$
denote the Fourier transformation with the measures
\begin{eqnarray}
d^4\eta_2\, e^{ik_2\cdot\eta_2}\,\,\, \text{and} \,\,\,
d^4\eta_1\, d^4 z\, e^{-ik_1\cdot\eta_1-i\ell\cdot z} ,
\end{eqnarray}
respectively, while ${\cal F}_1^{-1}$ and ${\cal F}_2^{-1}$ mark the inverse
Fourier transformation with the measures
\begin{eqnarray}
dy \, e^{i y\lambda}\,\,\, \text{and} \,\,\,
dx_1 dx_2 \, e^{i x_1\lambda_1+ i(x_2 - x_1)\lambda_2}.
\end{eqnarray}
We now implement the {\it factorization procedure} (see for instance \cite{Anikin:2009bf, Efremov:1984ip}) which contains the following steps:
(a) the decomposition of loop integration momenta around the corresponding dominant direction:
$k_i = x_i p + (k_i\cdot p)n + k_T$
within the certain light cone basis formed by the vectors $p$ and $n$ (in our case, $n^*$ and $n$);
(b) the replacement:
$d^4 k_i \Longrightarrow d^4 k_i \,dx_i \delta(x_i-k_i\cdot n)$
that introduces the fractions with the appropriated spectral properties;
(c) the decomposition of the corresponding propagator products around the dominant direction.
In Eqn.~(\ref{HadTen1-2}), we have (here, $x_{ij}=x_i-x_j$)
\begin{eqnarray}
\label{S-decom}
&&S(\ell-k_2) = S(x_{21}p_1-yp_2) +
\\
&&\frac{\partial S(\ell-k_2)}{\partial \ell_\rho} \Bigg|^{k_2=yp_2}_{\ell=x_{21}p_1} \, \ell^T_\rho + \ldots \,;
\nonumber
\end{eqnarray}
(d) the use of the collinear Ward identity:
\begin{eqnarray}
\frac{\partial S(k)}{\partial k_\rho} = S(k)\gamma_{\rho}S(k),\quad
S(k)=\frac{-\slashed k}{k^2 + i\varepsilon};
\nonumber
\end{eqnarray}
(e) performing of the Fierz decomposition for $\psi_\alpha (z) \, \bar\psi_\beta(0)$ in
the corresponding space up to the needed projections.

After factorization, the standard tensor, see Eqn.~(\ref{HadTen1-2}), is split into two terms: the first term includes
the correlator without the transverse derivative, while the second term contains the correlator with the transverse derivative,
see Eqns.~(\ref{S-decom}) and (\ref{ParFunB1})-(\ref{ParFunBperp}).

The non-standard contribution comes from the diagram depicted in Fig.~\ref{Fig-DY} (the right panel).
The corresponding hadron tensor takes the form \cite{AT-GP}:
\begin{eqnarray}
\label{HadTen2}
&&d{\cal W}_{(\text{Non-stand.})}^{\mu\nu}=
\\
&&
\int d^4 k_1\, d^4 k_2 \, \delta^{(4)}(k_1+k_2-q)
\text{tr}\big[
\gamma^\mu  {\cal F}(k_1) \gamma^\nu \bar\Phi(k_2)
\big]
\, ,
\nonumber
\end{eqnarray}
where the function ${\cal F}(k_1)$ reads
\begin{eqnarray}
\label{PhiF2}
&&{\cal F}(k_1)=
S(k_1) \gamma^\alpha \int d^4\eta_1\, e^{-ik_1\cdot\eta_1}\times
\nonumber\\
&&
\langle p_1, S^T | \bar\psi(\eta_1) \, gA_{\alpha}(0) \, \psi(0) |S^T,p_1\rangle \, .
\end{eqnarray}

For convenience, we introduce the unintegrated tensor $\overline{\cal W}_{\mu\nu}$ for the
factorized hadron tensor ${\cal W}_{\mu\nu}$ of the process. It reads
\begin{eqnarray}
\label{Uninteg-FacHadTen}
&&{\cal W}^{\mu\nu}=\int d^2 \vec{\textbf{q}}_T d{\cal W}^{\mu\nu}=\frac{2}{q^2}
\int d^2 \vec{\textbf{q}}_T \,\delta^{(2)}(\vec{\textbf{q}}_T) \times
\nonumber\\
&&
i\, \int dx_1 \, dy \,
\big[\delta(x_1/x_B-1) \delta(y/y_B-1)\big]
\overline{\cal W}^{\mu\nu}.
\end{eqnarray}
After calculation of all relevant traces in the factorized hadron tensor and after some algebra, we arrive at the following
contributions for the unintegrated hadron tensor (which involves all relevant contributions except the mirror ones):
the standard diagram depicted in Fig.~\ref{Fig-DY}, the left panel, gives us
\begin{eqnarray}
\label{DY-St}
&&\overline{\cal W}_{(\text{Stand.})}^{\mu\nu}
 + \overline{\cal W}_{(\text{Stand.},\,\partial_\perp)}^{\mu\nu}=\bar q(y)\,
\Bigg\{
\\
&&
- \frac{p_{1}^{\mu}}{y}\,
\varepsilon^{\nu S^T - p_2}\, \int dx_2 \frac{x_1-x_2}{x_1-x_2+i\epsilon} B^{(1)}(x_1,x_2)
\nonumber\\
&&  -
\Big[ \frac{p_{2}^{\nu}}{x_1} \varepsilon^{\mu S^T - p_2} + \frac{p_{2}^{\mu}}{x_1} \varepsilon^{\nu S^T - p_2} \Big]
x_1\int dx_2 \frac{B^{(2)}(x_1,x_2)}{x_1-x_2+i\epsilon}
\nonumber\\
&&+ \frac{p_{1}^{\mu}}{y} \,
\varepsilon^{\nu S^T - p_2}\, \int dx_2 \frac{B^{(\perp)}(x_1,x_2)}{x_1-x_2+i\epsilon}
\Bigg\}\,,
\nonumber
\end{eqnarray}
while the non-standard diagram presented in Fig.~\ref{Fig-DY}, the right panel,
contributes as
\begin{eqnarray}
\label{DY-NonSt}
&&\overline{\cal W}_{(\text{Non-stand.})}^{\mu\nu}=
\bar q(y)
\frac{p_{2}^{\mu}}{x_1}
\varepsilon^{\nu S^T -p_2}\times
\nonumber\\
&&
\int dx_2 \Big\{ B^{(1)}(x_1,x_2) +
B^{(2)}(x_1,x_2)
\Big\}.
\end{eqnarray}
Here we introduce the shorthand notation:
$\varepsilon^{A B C D}= \varepsilon^{\mu_1 \mu_2 \mu_3 \mu_4} A_{\mu_1} B_{\mu_2} C_{\mu_3} D_{\mu_4}$
with $\varepsilon^{0123}=1$.
Moreover, the parametrizing functions are associated with the following correlators:
\begin{eqnarray}
\label{ParFunB1}
&&i\varepsilon^{\alpha + S^T -} (p_1p_2)\, B^{(1)}(x_1,x_2)=
\\
&&{\cal F}_2\Big[\langle p_1, S^T| \bar\psi(\eta_1)\, \gamma^+ \, gA^\alpha_\perp(z)\, \psi(0) | S^T,p_1 \rangle \Big]\,,
\nonumber\
\end{eqnarray}
\begin{eqnarray}
\label{ParFunB2}
&&i\varepsilon^{+ \beta S^T -} (p_1p_2)\, B^{(2)}(x_1,x_2)=
\\
&&{\cal F}_2\Big[\langle p_1, S^T| \bar\psi(\eta_1)\, \gamma^\beta_\perp \, gA^+(z)\, \psi(0) | S^T,p_1 \rangle \Big]\,,
\nonumber\
\end{eqnarray}
\begin{eqnarray}
\label{ParFunBperp}
&& i p_1^+ \varepsilon^{\rho + S^T -} (p_1p_2) B^{(\perp)}(x_1,x_2)=
\\
&&{\cal F}_2\Big[\langle p_1, S^T| \bar\psi(\eta_1)\, \gamma^+\, \big(\partial^\rho_\perp\, gA^+(z)\big)\, \psi(0) | S^T,p_1 \rangle \Big]\,,
\nonumber
\end{eqnarray}
where $\eta_1=\lambda_1\tilde n$, $z=\lambda_2 \tilde n$, and
the light-cone vector $\tilde n$ is a dimensionful analog of $n$ ($\tilde n^-=p_2^-/(p_1p_2)$).

As known from \cite{AT-GP}, the function $B^{(1)}(x_1,x_2)$ for the DY process
can be unambiguously written as
\begin{eqnarray}
\label{B1-fun}
B^{(1)}(x_1,x_2)=\frac{T(x_1,x_2)}{x_1-x_2+i\varepsilon}\,,
\end{eqnarray}
where the function $T(x_1,x_2)\in\Re\text{e}$
parametrizes the corresponding projection of $\langle \bar\psi\, G_{\alpha\beta}\,\psi \rangle$, {\it i.e.}
\begin{eqnarray}
\label{parT}
&&\varepsilon^{\alpha + S^T -}\,(p_1p_2)\, T(x_1,x_2)=
\\
&&{\cal F}_2\Big[ \langle p_1, S^T | \bar\psi(\eta_1)\, \gamma^+ \,
\tilde n_\nu G^{\nu\alpha}_T(z) \,\psi(0)
|S^T, p_1 \rangle\Big]\,.
\nonumber
\end{eqnarray}
Notice that we have derived (see \cite{AT-GP}) the certain complex prescription in the
{\it  r.h.s.} of (\ref{B1-fun}) within the contour gauge. In this letter, we assume that
the same prescription takes place in the Feynman gauge too \footnote{Generally speaking, in the Feynman gauge
the arguments how to derive the certain complex prescription differ from that we used in \cite{AT-GP}.
For example, the prescription can be defined by ordering of operator positions on the light-cone direction.}.
With respect to the functions $B^{(2)}(x_1,x_2)$ and  $B^{(\perp)}(x_1,x_2)$, we  demonstrate below that
these functions do not possess the gluon poles and, therefore, cannot be presented in the form of (\ref{B1-fun}).

Summing up all contributions from the standard and non-standard diagrams, we finally obtain
the expression for the unintegrated hadron tensor as
\nopagebreak
\begin{widetext}
\begin{eqnarray}
\label{DY-ht-1}
&&\overline{\cal W}^{\mu\nu}=
\overline{\cal W}_{(\text{Stand.})}^{\mu\nu} + \overline{\cal W}_{(\text{Stand.},\,\partial_\perp)}^{\mu\nu}+
\overline{\cal W}_{(\text{Non-stand.})}^{\mu\nu}=
\bar q(y)\,
\Bigg\{ \Big[ \frac{p_{2}^{\mu}}{x_1} - \frac{p_{1}^{\mu}}{y} \Big] \,
\varepsilon^{\nu S^T -p_2}\, \int dx_2 B^{(1)}(x_1,x_2) +
\nonumber\\
&& \frac{p_{2}^{\mu}}{x_1} \,
\varepsilon^{\nu S^T - p_2}\, \int dx_2 B^{(2)}(x_1,x_2) -
\Big[ \frac{p_{2}^{\nu}}{x_1} \varepsilon^{\mu S^T - p_2} + \frac{p_{2}^{\mu}}{x_1} \varepsilon^{\nu S^T - p_2} \Big]
x_1\int dx_2 \frac{B^{(2)}(x_1,x_2)}{x_1-x_2+i\epsilon} +
\nonumber\\
&& \frac{p_{1}^{\mu}}{y} \,
\varepsilon^{\nu S^T - p_2}\, \int dx_2 \frac{B^{(\perp)}(x_1,x_2)}{x_1-x_2+i\epsilon}
\Bigg\}\,,
\end{eqnarray}
\end{widetext}
Notice that the first term in Eqn.~(\ref{DY-ht-1}) coincides with the hadron tensor calculated within the light-cone gauge
$A^+=0$.

\section{QED gauge invariance of hadron tensor}

Let us now discuss the QED gauge invariance of the hadron tensor.
From Eqn.~(\ref{DY-ht-1}),
we can see that the QED gauge invariant combination is
\begin{eqnarray}
\label{GI-comb}
&&{\cal T}^{\mu\nu}=\Big[ \frac{p_{2}^{\mu}}{x_1} - \frac{p_{1}^{\mu}}{y} \Big] \,
\varepsilon^{\nu S^T -p_2},\,
\nonumber\\
&&\text{with}\quad
q_\mu {\cal T}^{\mu\nu} = q_\nu {\cal T}^{\mu\nu}=0.
\end{eqnarray}
We can see that there is a single term with $p_{2}^{\nu}$ which does not have
a counterpart to construct the gauge-invariant combination
\begin{eqnarray}
\label{GI-com-2}
\frac{p_{2}^{\mu}}{x_1} - \frac{p_{1}^{\mu}}{y}.
\end{eqnarray}
Therefore, the second term in  Eqn.~(\ref{DY-St}) should be equal to zero.
This also leads to nullification of the second term in Eqn.~(\ref{DY-NonSt}).

Hence, the only way to get the QED gauge invariant combination (see (\ref{GI-comb})) is
to combine the first terms in Eqns.~(\ref{DY-St}) and (\ref{DY-NonSt}). This combination justifies the treatment of gluon pole in $B^{(1)}(x_1,x_2)$
using the complex prescription.

In addition, we conclude that the third term in (\ref{DY-St}) does not contribute to SSA.

\begin{figure*}[ht]
\centerline{\includegraphics[width=0.3\textwidth]{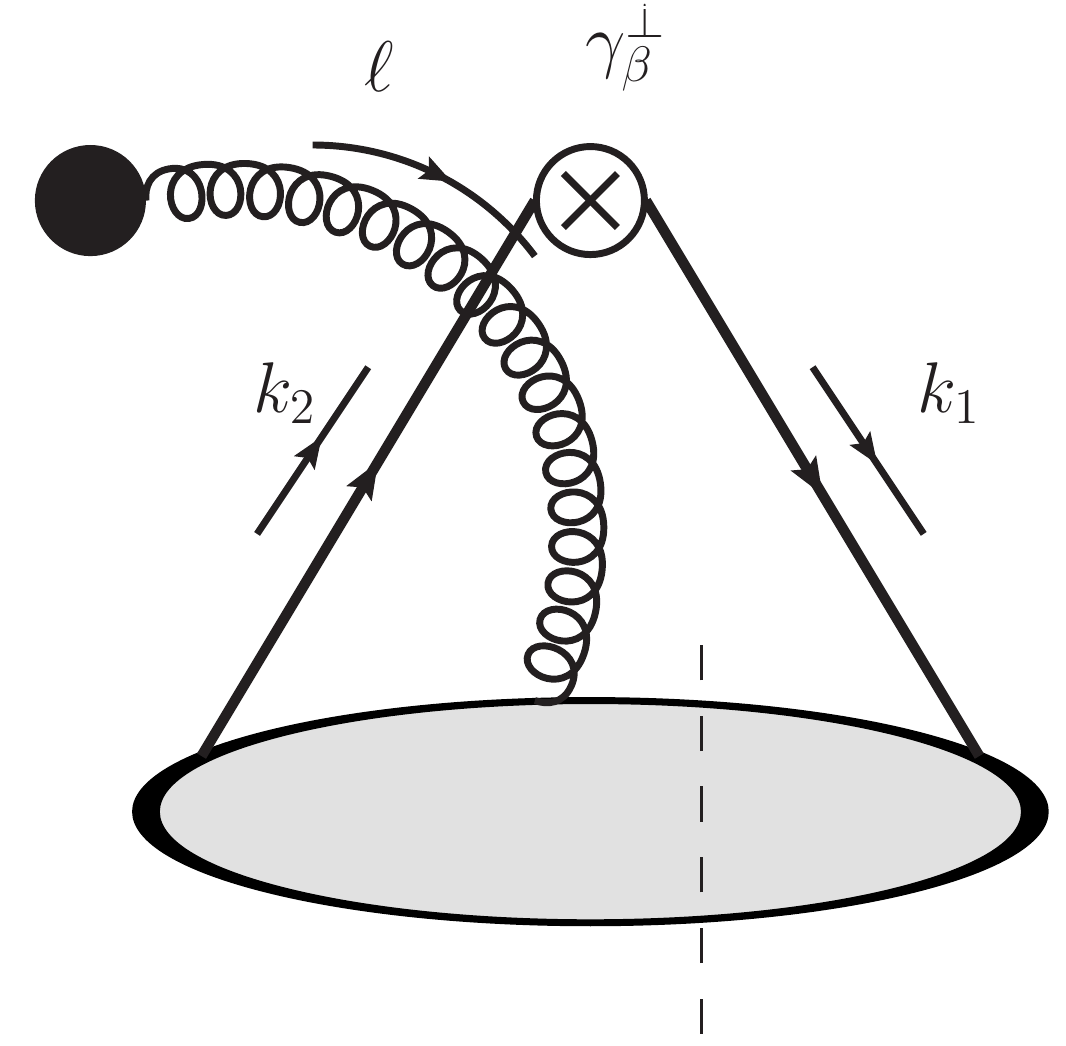}
\hspace{2.cm}\includegraphics[width=0.26\textwidth]{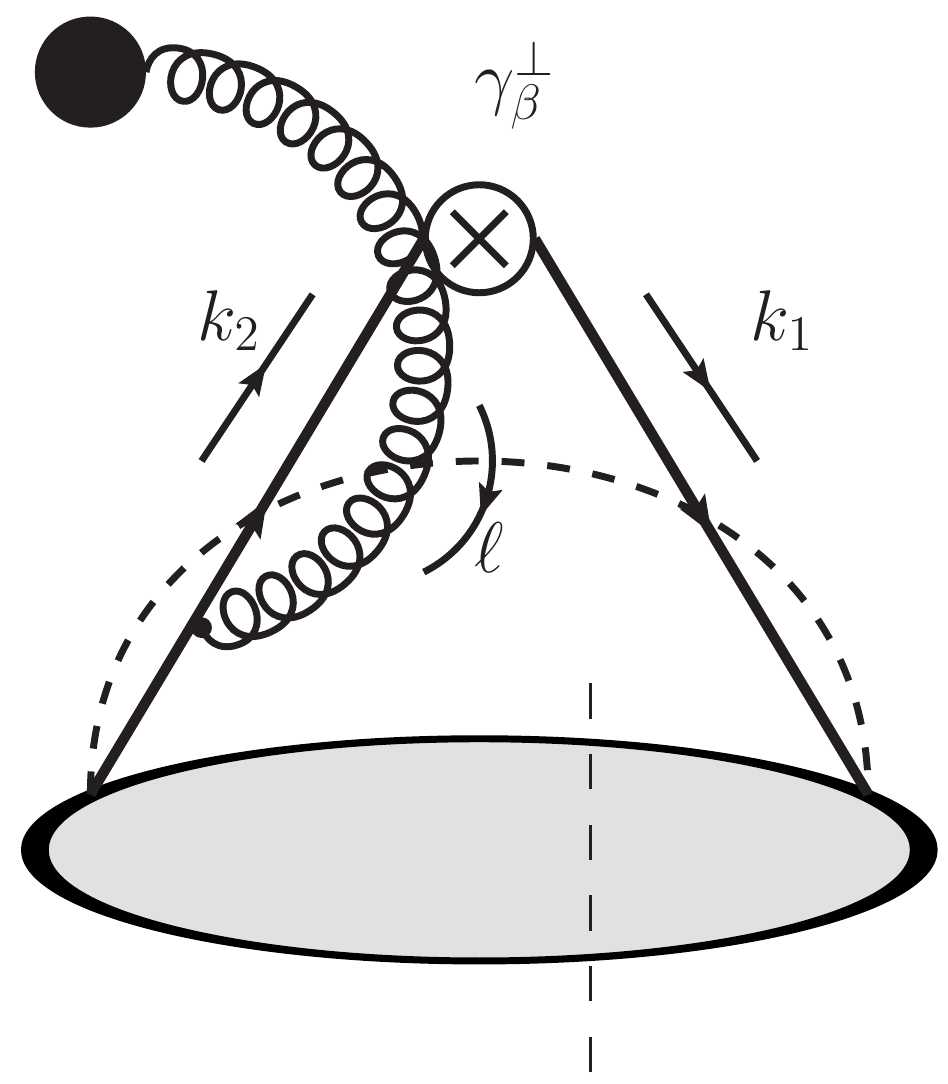}}
\vspace{1cm}
\caption{The matrix element (correlator) of nonlocal twist-3 quark-gluon operator within the momentum
representation. Here $\ell=k_2-k_1$ and
$k_1=(x_1 p^+_1, k^-_1, \vec{{\bf k}}_{1\,\perp})$, $k_2=(x_2 p^+_1, k^-_2, \vec{{\bf k}}_{2\,\perp})$}
\label{Fig-DY-2}
\end{figure*}

The suggested proof explores only the gauge and Lorentz invariance.
Let us consider the other reasoning to justify these properties of correlators,
starting with the correlator which generates the function $B^{(2)}(x_1,x_2)$:
\begin{eqnarray}
\label{Corr-1}
&&\int (d\lambda_1\, d\lambda_2) e^{-ix_1\lambda_1 - i(x_2-x_1)\lambda_2}\times
\nonumber\\
&&\langle p_1, S^T| \bar\psi(\lambda_1 \tilde n)\, \gamma_\perp^\beta \, A^+(\lambda_2 \tilde n)\, \psi(0) | S^T,p_1 \rangle =
\nonumber\\
&&
i\varepsilon^{+ \beta S^T - }\, (p_1 p_2)\, B^{(2)}(x_1,x_2)\,.
\end{eqnarray}
We are going over to the momentum representation for the correlator from the l.h.s. of Eqn.~(\ref{Corr-1}).
Schematically, we have
\begin{eqnarray}
\label{Corr-2}
\Big[ \bar u(k_1) \gamma^\perp_\beta u(k_2)\Big] \times  .... \times \frac{1}{\ell^2 + i\varepsilon}\,,
\end{eqnarray}
where the gluon momentum is $\ell = k_2-k_1$ and $k_1=(x_1 p^+_1, k^-_1, \vec{{\bf k}}_{1\,\perp})$, $k_2=(x_2 p^+_1, k^-_2, \vec{{\bf k}}_{2\,\perp})$.
This situation has been illustrated in Fig.~\ref{Fig-DY-2}, see the left panel.
Up to the order of $g$, we are also able to write down that (see Fig.~\ref{Fig-DY-2}, the right panel)
\begin{eqnarray}
\label{Corr-3}
\Big[ \bar u(k_1) \gamma^\perp_\beta {\cal S}(k_2) u(k_1 )\Big] \times  .... \times \frac{1}{\ell^2 + i\varepsilon}\,,
\end{eqnarray}
where ${\cal S}(k_2)=S(k_2)\gamma^+$.
From both these equations, it is clear that to get the non-zero contribution we must have either $\vec{{\bf k}}_{1\,\perp}\neq 0$
or $\vec{{\bf k}}_{2\,\perp}\neq 0$. Indeed,
\begin{eqnarray}
\Big[ \bar u(k_1) \gamma^\perp_\beta {\cal S}(k_2) u(k_1 )\Big] \Rightarrow S_{\beta k_2 + k_1}= k^\perp_{2\,\beta} k^+_1 + k^\perp_{1\,\beta} k^+_2\,.
\end{eqnarray}
Therefore, the gluon propagator in Eqns.~(\ref{Corr-2}) and (\ref{Corr-3}) takes the following form (cf. \cite{Braun}):
\begin{eqnarray}
\label{gluon-prop}
\frac{1}{\ell^2+i\varepsilon}= \frac{1}{2(x_2-x_1)p^+_1 \ell^- - \vec{{\bf l}}^2_{\perp} +i\varepsilon}.
\end{eqnarray}
One can conclude that, in the case of the substantial transverse component of the momentum,
there are no sources for the gluon poles at $x_1=x_2$. As a result, the function  $B^{(2)}(x_1,x_2)$ has no gluon poles and,
due to T-invariance \cite{Efremov:1984ip} ($B^{(2)}(x_1,x_2)= - B^{(2)}(x_2,x_1)$), obeys $B^{(2)}(x,x) = 0$.

On the other hand, if we have $\gamma^+$ in the correlator (see Eqn. (\ref{ParFunB1})), the transverse components of gluon momentum
are not substantial and can be neglected. That ensures the existence of the gluon poles for the function $B^{(1)}(x_1,x_2)$.
This corresponds to the fact that the Sivers function, being related to gluon poles, contains the "leading twist" projector $\gamma^+$.
Moreover, we may conclude that the structure $\gamma^+ (\partial^\perp A^+)$ does not produce the imaginary part as well as SSA in the Feynman gauge.

\section{Conclusions and discussions}

Working within the Feynman gauge, we derive the
QED gauge invariant (unintegrated) hadron tensor for the polarized DY process:
\begin{eqnarray}
\label{DY-ht-GI}
&&\overline{\cal W}_{\text{GI}}^{\mu\nu}=
\overline{\cal W}_{(\text{Non-stand.})}^{\mu\nu} +
\overline{\cal W}_{(\text{Stand.})}^{\mu\nu}=
\nonumber\\
&&\bar q(y)
\Big[ \frac{p_{2}^{\mu}}{x_1} - \frac{p_{1}^{\mu}}{y} \Big]
\varepsilon^{\nu S^T -p_2} \hspace{-1.5mm}\int dx_2 B^{(1)}(x_1,x_2).
\end{eqnarray}
After calculating the imaginary part (or, in other words, after adding the mirror contributions),
and, then integrating over $x_1$ and $y$
(see Eqn.~(\ref{Uninteg-FacHadTen})), we get
the QED gauge invariant hadron tensor as
\begin{eqnarray}
\label{DY-ht-GI-2}
W_{\text{GI}}^{\mu\nu}= \bar q(y_B)\,
\Big[ \frac{p_{2}^{\mu}}{x_B} - \frac{p_{1}^{\mu}}{y_B} \Big] \,
\varepsilon^{\nu S^T -p_2}\,  T(x_B ,x_B)\,.
\end{eqnarray}
This expression fully coincides with the hadron tensor which
has been derived within the light-cone gauge for gluons.

\begin{figure*}[ht]
\centerline{\includegraphics[width=0.45\textwidth]{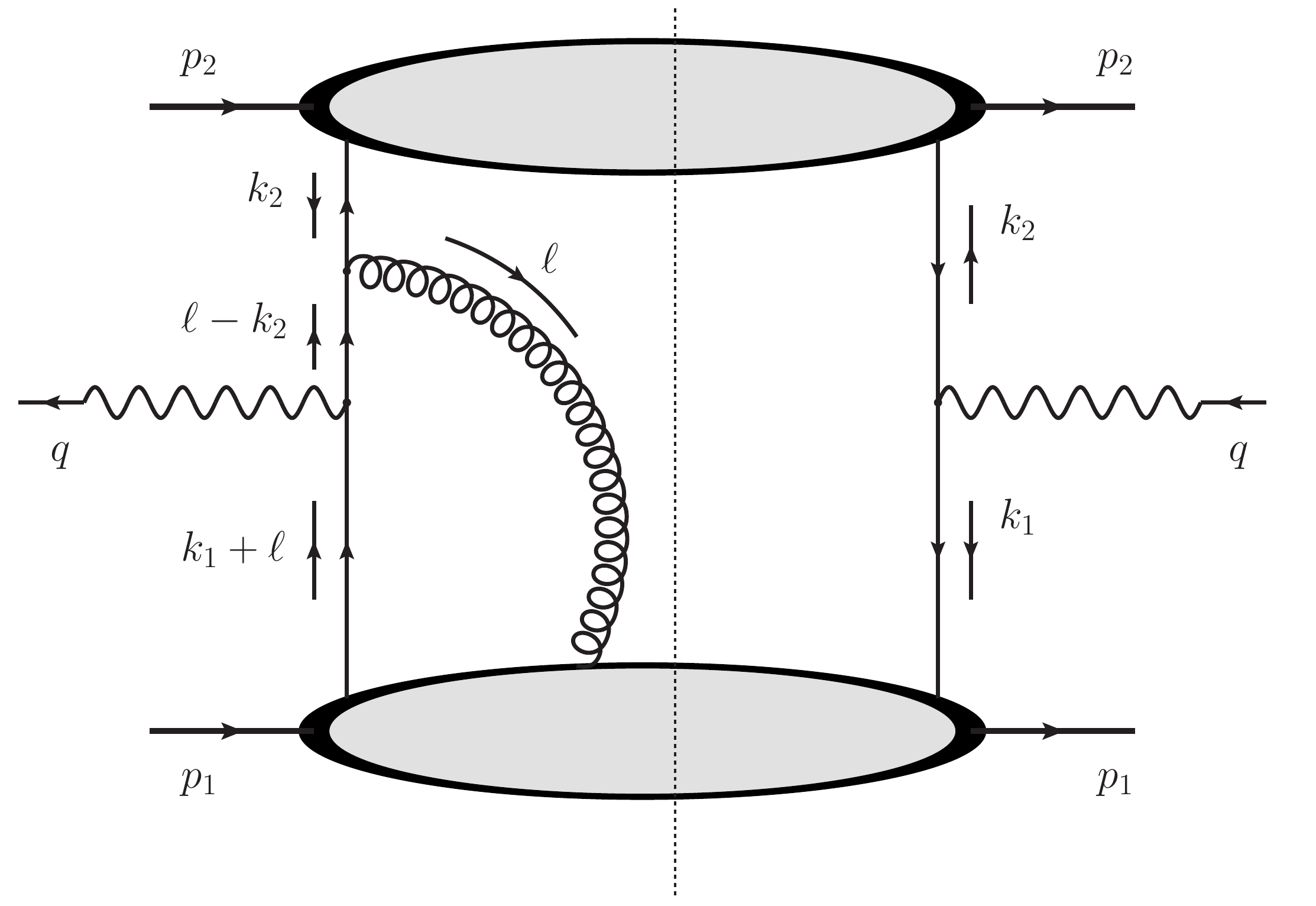}
\hspace{1.cm}\includegraphics[width=0.45\textwidth]{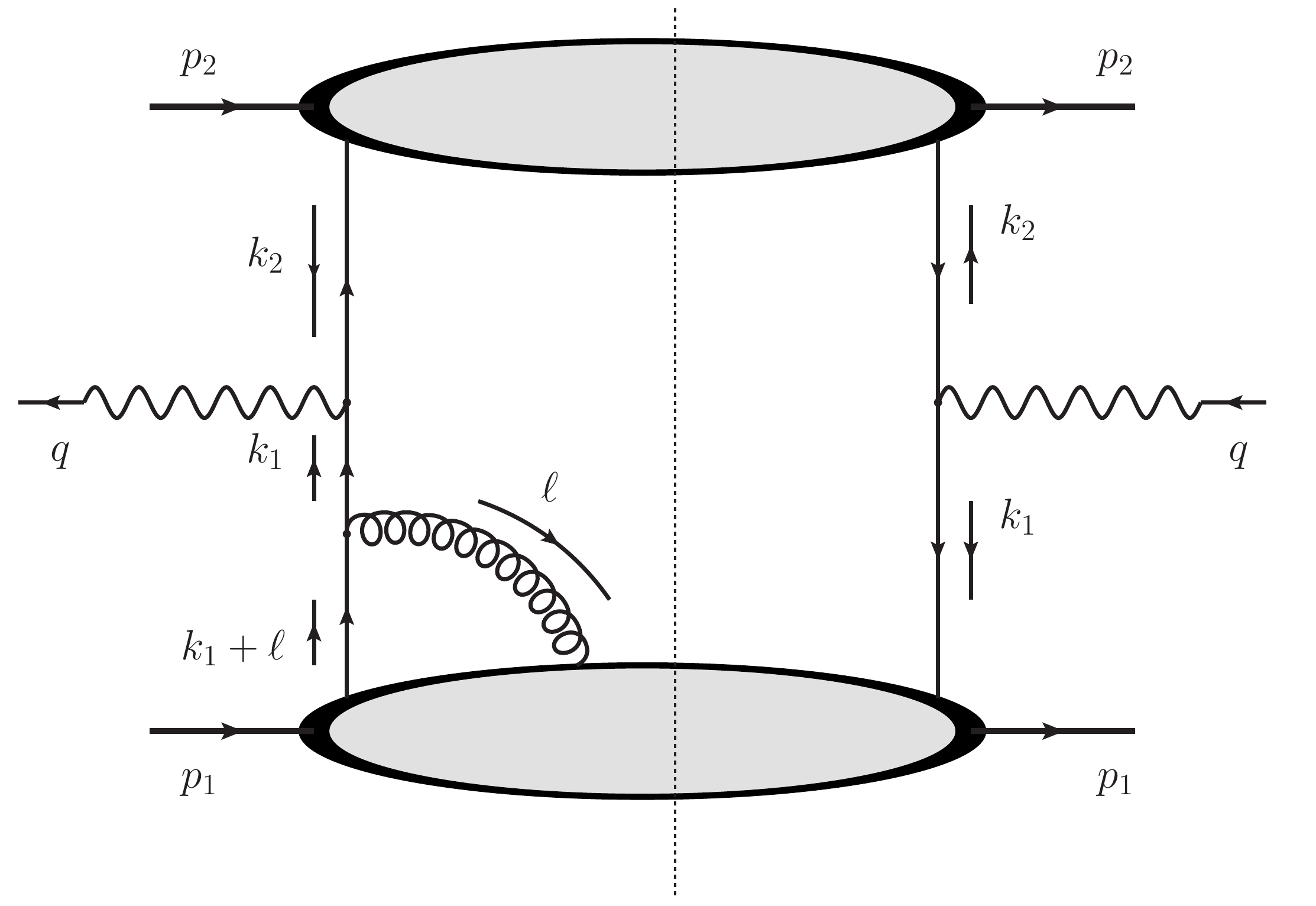}}
  \caption{The Feynman diagrams which contribute to the polarized Drell-Yan hadron tensor.}
\label{Fig-DY}
\end{figure*}

Moreover, the factor of $2$ in the hadron tensor that
we found within the axial-type gauge \cite{AT-GP} is still present in the frame
of the Feynman gauge.
In order to show this factor of $2$, let us introduce the mutually orthogonal basis (see \cite{Barone})
as
\begin{eqnarray}
\label{vecZ}
Z_\mu= \widehat p_{1\,\mu} - \widehat p_{2\,\mu} \equiv x_B \, p_{1\,\mu}- y_B\, p_{2\,\mu} \,
\end{eqnarray}
and
\begin{eqnarray}
\label{vecX-Y}
&&\hspace{-0.5cm}X_\mu= -\frac{2}{s} \biggl[
(Z p_2)\biggl(p_{1\, \mu} - \frac{q_\mu}{2x_B} \biggr) -
(Z p_1)\biggl(p_{2\, \mu} - \frac{q_\mu}{2y_B} \biggr)
\biggr],
\nonumber\\
&&\hspace{-0.5cm}Y_\mu=\frac{2}{s} \, \varepsilon_{\mu p_1 p_2 q}.
\end{eqnarray}
Here $\widehat p_{i\,\mu}$ are the partonic momenta
($q^\mu=\widehat p_{1\,\mu}+\widehat p_{2\,\mu}$).
With the help of (\ref{vecZ}) and (\ref{vecX-Y}), the lepton momenta can be written as
(this is the lepton c.m. system)
\begin{eqnarray}
\label{lepmom}
&&l_{1\,\mu} = \frac{1}{2} q_\mu  + \frac{Q}{2} f_\mu(\theta,\varphi; \hat X, \hat Y, \hat Z)\, ,
\nonumber\\
&&l_{2\,\mu} = \frac{1}{2} q_\mu  - \frac{Q}{2} f_\mu(\theta,\varphi; \hat X, \hat Y, \hat Z)\, ,
\end{eqnarray}
where $\hat A = A/\sqrt{-A^2}$ and
\begin{eqnarray}
&&f_\mu(\theta,\varphi; \hat X, \hat Y, \hat Z)=
\\
&&\hat X_\mu\, \cos\varphi \,\sin\theta +
\hat Y_\mu\, \sin\varphi \,\sin\theta + \hat Z_\mu\, \cos\theta \, .
\nonumber
\end{eqnarray}
Within this frame, the contraction of the lepton tensor with the gauge invariant
hadron tensor (\ref{DY-ht-GI-2}) reads
\begin{eqnarray}
\label{Contraction}
{\cal L}_{\mu\nu} \, W_{\text{GI}}^{\mu\nu} =
-2 \cos\theta\, \varepsilon^{l_1 S^T p_1 p_2} \,\bar q(y_B)\, T(x_B,x_B)\, .
\end{eqnarray}
We want to emphasize that this expression in (\ref{Contraction}) differs by the factor of $2$ in comparison with the case where
only one diagram (presented in Fig. \ref{Fig-DY}, the left panel) has been included in the (gauge non-invariant)
hadron tensor, {\it i.e.}
\begin{eqnarray}
\label{Diff2}
{\cal L}_{\mu\nu} \, W_{(\text{Stand.})}^{\mu\nu} =
\frac{1}{2}\, {\cal L}_{\mu\nu} \, W_{\text{GI}}^{\mu\nu} \, .
\end{eqnarray}
Therefore, from the practical point of view, if we neglect the diagram in
Fig. \ref{Fig-DY} (right panel) or, in other words, if we
use the QED gauge non-invariant hadron tensor, it yields the error of the factor of two.

Further, based on the light-cone dynamics we argue that there are no gluon poles
in the correlators $\langle\bar\psi\gamma_\perp A^+\psi\rangle$.
This means that the function $B^{(2)}(x_1,x_2)$ does not have the representation similar to (\ref{B1-fun}).
We also show that
the Lorentz and QED gauge invariances of the hadron tensor calculated within the Feynman gauge
require that the function $B^{(2)}(x_1,x_2)$ cannot have gluon poles.

The fact that the function $B^{(2)}(x_1,x_2)$ cannot be presented in the form of (\ref{B1-fun})
directly leads to the absence of $dT/dx$ in the final expression of the gauge-invariant hadron tensor.
Indeed, from (\ref{DY-St}), one can see that $B^{(2)}(x_1,x_2)$ contributes to the standard hadron tensor as
\begin{eqnarray}
\label{DY-St-dT}
\Big[ p_{2}^{\nu} \varepsilon^{\mu S^T - p_2} + p_{2}^{\mu} \varepsilon^{\nu S^T - p_2} \Big]
\int dx_2 \frac{B^{(2)}(x_1,x_2)}{x_1-x_2+i\epsilon}.
\end{eqnarray}
In order to obtain the $dT/dx$-contribution,  we have to impose the representation (\ref{B1-fun})
on $B^{(2)}(x_1,x_2)$ and, then perform the integration over $dx_2$ by part.
However, as shown above, $B^{(2)}(x_1,x_2)$ does not have the representation (\ref{B1-fun}).

This property seems to be natural from the point of view of gluon poles relation \cite{Boer:2003cm} to Sivers functions
as the latter is related to the projection $\gamma^+$. As for the function  $B^{(\perp)}(x_1,x_2)$,  the transverse derivative
of Sivers function resulting from taking its moments may act on both integrand and boundary value. Our result suggests that only the
action on the boundary value  related to  $B^{(1)}(x_1,x_2)$ should produce SSA.
It is certainly not unnatural keeping in mind that the integrand
differentiation is present even for simple straight-line contours which are not producing SSA.

\nopagebreak

\section*{Acknowledgments}

We thank A.V.~Efremov and A.~Prokudin for useful discussions.
The work by I.V.A. was partially supported by the Heisenberg-Landau Program of the
German Research Foundation (DFG).

\end{document}